\documentclass[prc,twocolumn,groupedaddress,floatfix]{revtex4}
\usepackage{graphicx}
\usepackage{dcolumn}
\usepackage{bm}

\begin{document}

\title{Comment on "Pygmy dipole response of proton-rich argon nuclei in
random-phase approximation and no-core shell model"}
\author{N. Paar\footnote{npaar@phy.hr}}
\affiliation{Physics Department, Faculty of Science, University of Zagreb, 
Croatia}
\date{\today}

 

\maketitle
In a recent article by C. Barbieri, E. Caurier, K. Langanke, and
G. Mart\'inez-Pinedo~\cite{Bar.08}, low-energy
dipole excitations were studied in proton-rich $^{32,34}$Ar 
with random-phase approximation (RPA) 
and no-core shell model (NCSM) using correlated 
realistic nucleon-nucleon interactions obtained by the
unitary correlation operator method (UCOM)~\cite{Fel.98}.
The main objective of this Comment is to argue that the article~\cite{Bar.08}
contains an inconsistency with respect to previous
study of excitations in the same UCOM-RPA framework using 
identical correlated Argonne V18 interaction~\cite{Paa.06},
it does not provide any evidence that the low-lying state declared as
pygmy dipole resonance in $^{32}$Ar  indeed has the resonance-like 
structure, and that prior to studying
exotic modes of excitation away from the valley of stability
one should ensure that the model provides reliable 
description of available experimental data on nuclear
ground state properties and excitations in nuclei.
Although the authors aimed at testing the UCOM based theory 
at the proton drip line, available experimental data that are used 
as standard initial tests of theory frameworks at the proton drip line
have not been considered in the UCOM case (e.g., binding energies,
one-proton separation energies, two-proton separation energies).

The parametrized correlation functions that describe 
the short-range correlations in the UCOM 
framework have been constrained at the level of the 
two-nucleon system, and their ranges are determined 
by fitting the NCSM binding energies  of $^{3}$H and 
$^{4}$He to the experimental values~\cite{Rot.07}. 
In this way, some of the missing ingredients 
(e.g., the three-body force, higher orders in cluster 
expansion) are effectively taken into account and
included in the parameters of correlation functions.
Despite the fact that the short-range properties of the
correlated realistic nucleon-nucleon interaction are well
behaved and some of the three-body effects are implicitly 
included, the UCOM framework cannot provide quantitative
description of nuclear structure properties within
Hartree-Fock (HF) and Fermionic molecular 
dynamics (FMD).
Therein the binding energies of nuclei across the
nuclide chart are dramatically 
underestimated (in HF up to 50\%), and the radii are smaller
by 1-3 fm  when compared to experimental 
data~\cite{Rot.06}. A considerable portion of the missing
correlations can be recovered, e.g., by many-body 
perturbation theory on top of HF~\cite{Rot.06}, or by
introducing additional purely phenomenological terms with
new free parameters supplemented to the 
correlated realistic nucleon-nucleon interaction
(e.g. in FMD~\cite{Che.07}).  

In Ref.~\cite{Paa.06} the UCOM HF + RPA based
on correlated Argonne V18 interaction is
introduced and tested on several nuclei from $^{16}$O 
toward $^{208}$Pb. It has been pointed out
that due to unrealistic
descriptions of the HF ground state (e.g., single-particle
spectra are extremely wide), the missing long-range
correlations and three-body interaction, 
the energies of giant resonances
are strongly overestimated in comparison
to experimental data and previous studies. In particular, 
excitation energies of isovector giant dipole resonances (IVGDR)
are $\approx$3-8 MeV higher than those of
the experimental values~\cite{Paa.06}. Despite these facts,
C. Barbieri, E. Caurier, K. Langanke, and
G. Mart\'inez-Pinedo employ the same UCOM HF + RPA
based on identical correlated Argonne V18 interaction
in the study of an exotic excitation in the $1^-$ channel:
proton pygmy dipole resonance (PPDR) in $^{32,34}$Ar~\cite{Bar.08}. 
One should be concerned by the fact that UCOM RPA 
which seriously overestimates IVGDR across the nuclide chart
from $^{16}$O toward $^{208}$Pb\cite{Paa.06}, in the
case of Ar isotopes appears in rather good
agreement with empirical estimates~\cite{Bar.08} and
relativistic QRPA which on the other hand 
quantitatively also describe giant resonances in other
nuclei~\cite{Paa.05}. 
The reported UCOM RPA results for $^{32,34}$Ar are 
obviously inconsistent with previous systematic
study based on the same model and effective interaction,
employed along the nuclide chart including the region of 
medium heavy nuclei (Fig. 10 in Ref.~\cite{Paa.06}). 

Although in Ref.~\cite{Bar.08} Barbieri et al. claim
that a clear peak associated with pygmy dipole 
resonance has been found in UCOM RPA and NCSM
calculations in $^{32}$Ar, they do not provide any
supporting evidence on the resonance-like structure 
of the corresponding state.
Based on transition densities only, one cannot
draw conclusions on the collectivity of the
low-energy excitations; i.e., the same transition 
densities as presented in Ref.~\cite{Bar.08}  in 
the upper panel of Fig. 3, could also have the origin 
only in single-particle transitions~\cite{Vre.01}. In 
Ref.~\cite{Paa.05} it has been pointed out that 
actually the pairing correlations play an essential
role in building up the collectivity of
the PPDR mode in $^{32}$Ar and other medium heavy
nuclei close to the proton drip line. 
This means that the structure of a single relevant low-energy 
state becomes more distributed; i.e., a considerable number
of two-quasiparticle configurations contribute to a particular
excitation mode. In the limiting case
when pairing correlations are absent, the collectivity of 
the low-lying states vanishes; i.e., each low-lying state is
dominated mainly
by a single-particle transition. In the article by Barbieri et al.
the pairing correlations are not properly included
in HF+RPA calculations
(partial occupation of $\pi 0d_{3/2}$ orbit is set to 1/2, other
orbits are fully occupied or empty; there is no pairing interaction in
ground state calculations and residual RPA interaction).
In contrast, shell model calculations, performed in the
full $0\hbar \omega$ space, include pairing with all the
possible seniorities. Unfortunately, it remains unknown whether
there is some difference in the nature of the observed 
low-energy excitations in HF + RPA and shell model calculations.
Although the issue of collectivity in the low-lying dipole
states is simply ignored,  the term "pygmy dipole resonance"
has extensively been used throughout the article~\cite{Bar.08}.

Recent study of giant resonances using correlated 
realistic nucleon-nucleon interactions which include couplings
with complex configurations (2p2h), indicates
that actually the UCOM RPA response
of giant resonances  considerably shifts toward
lower energies~\cite{Pap.07}. In the case of dipole excitations
in $^{16}$O and $^{40}$Ca, the transition strengths are
systematically lowered by
$\approx$ 8 MeV. The same effect should also appear in
the case of $^{32}$Ar; i.e., one could expect that 
the UCOM RPA centroid energy of low-lying states will
be shifted from the value of 9.15 MeV reported in Ref.~\cite{Bar.08} 
to significantly lower values. It remains a puzzle why the NCSM
calculations for $^{32,34}$Ar~\cite{Bar.08} result
in strength distributions very similar to those of UCOM RPA (the centroid
energies are different by less than 1 MeV), although
the former includes up to 2p2h configuration spaces. 
Prior to implementation of UCOM NCSM on
excitations in exotic nuclei, the effect of the 2p2h model
space should have carefully been checked on stable
nuclei (e.g. $^{16}$O, $^{40}$Ca, $^{48}$Ca, etc.)
where extensive experimental data on giant 
resonances already exist, as has been done, e.g., 
in the UCOM second-RPA study~\cite{Pap.07}.

Finally, planned experiments aimed at observing the PPDR mode in
Ar isotopes do not present the best choice for testing the accuracy
of UCOM RPA and NCSM approaches. Prior to studying exotic modes
of excitation, theory models should be tested on the large set of already
existing experimental data on the nuclear ground state and excitation
properties in nuclei. In addition, there is a whole set of data available
on the properties of nuclei at proton drip line (e.g., binding energies,
one-proton separation energies, two-proton separation energies) 
that should be used for testing the accuracy of UCOM based
approaches at the proton drip line before studying excitations. 
Finally, there is a considerable amount of data on giant resonances
in stable nuclei, as well as recent data on pygmy dipole resonances
in the unstable nucleus $^{132}$Sn~\cite{Adr.05}.  All these data have
systematically been used in numerous studies and should be 
used in the first place to validate the implementation of effective
interactions in nuclear 
many-body theories. Over the past decades, much effort 
has been expended to achieve the microscopic description of
nuclear ground state and excitations, and available experimental
data provide a crucial test of nuclear structure models.
These essential studies should not be bypassed in the
case of UCOM based theory if the same approach aims at
description of exotic modes of excitation in nuclei away from
the valley of stability.

\acknowledgments
This work was supported by the Unity through Knowledge
Fund (UKF Grant No. 17/08).

\end{document}